\documentstyle[12pt,aasms4]{article}

\renewcommand{\thefootnote}{\fnsymbol{footnote}}

\slugcomment{Published in: {\bf Chinese Physics Letters 16 (1999) 775}}

\lefthead{HUANG Yong-feng {\sl et al.} }
\righthead{HUANG Yong-feng {\sl et al.} }

\begin{document}

\title{Overall Evolution of Realistic Gamma-Ray Burst Remnant \\ 
       and Its Afterglow\footnote{Supported by the National Natural 
       Science Foundation of China under Grant Nos. 19773007 and 
       19825109, and the National Climbing Project on Fundamental Researches.} }

\renewcommand{\thefootnote}{\arabic{footnote}}

\author{HUANG Yong-feng (\hspace{1.7cm})\altaffilmark{1}, \hspace{1mm} 
        DAI Zi-gao (\hspace{1.7cm})\altaffilmark{1}, \hspace{1mm}	
        Lu Tan (\hspace{1.2cm})\altaffilmark{1,2}}

\altaffiltext{1}{\sl Department of Astronomy, Nanjing University, 
		     Nanjing 210093} 

\altaffiltext{2}{\sl Laboratory of Cosmic Ray and High Energy Astrophysics, 
		     Institute for high Energy physics, 
		     Chinese Academy of Sciences, Beijing 100039}

\begin{abstract}

Conventional dynamic model of gamma-ray burst remnants is found to be
incorrect for adiabatic blastwaves during the non-relativistic 
phase. A new model is derived, which is shown to be correct 
for both radiative and adiabatic blastwaves during both 
ultra-relativistic and non-relativistic phase. Our model also 
takes the evolution of the radiative efficiency into account. 
The importance of the transition from the ultra-relativistic 
phase to the non-relativistic phase is stressed.

\vspace{0.3cm}

\noindent
{\sl PACS: 95.30.Lz, 98.70.Rz, 97.60.Jd} 

\end{abstract}

\vspace{0.5cm}

The origin of gamma-ray bursts (GRBs) has remained unknown for 
over 30 years.$^{1,2}$ A major breakthrough appeared in early 1997, 
when the Italian-Dutch BeppoSAX satellite observed X-ray afterglows 
from GRB 970228 for the first time.$^3$ Since then, X-ray afterglows 
have been observed from about 15 GRBs, of which ten events were detected 
optically and five bursts were also detected in radio wavelengths. 
The cosmological origin 
of at least some GRBs is firmly established. The so called 
fireball model\,$^{4,5}$ is strongly favoured, which is found 
successful at explaining the major features of the low energy 
light curves.$^{6 - 9}$ 

In the fireball model, low energy afterglows are generated by 
ultra-relativistic fireballs, which first give birth to GRBs 
through internal or external shock waves and then decelerate 
continuously due to collisions with the interstellar medium (ISM). 
The dynamics of the expansion has been investigated
extensively.$^{6 - 9}$ Both analytic solutions and numerical 
approaches are available. It is a general conception that 
current models describe the gross features of the process very
well and further improvements are possible only by considering 
some details. However, we find that three serious problems are 
associated with the popular model.

First, it is usually assumed that the expansion is  
ultra-relativistic. Then for an adiabatic fireball, the 
evolution of the bulk Lorentz factor is derived to be: 
\begin{equation}
\gamma \approx (200 - 400) E_{51}^{1/8} n_{1}^{-1/8} t^{-3/8},
\end{equation}
where $E_{51} = E_0 / (10^{51} {\rm erg})$ with $E_0$ the initial
fireball energy,  
$n_1 = n / (1 {\rm cm}^{-3})$ with $n$ the ISM number density, 
and $t$ is  
observer's time in unit of s.$^{6 - 9}$ The radius of the 
blastwave scales as $R \propto t^{1/4}$. Based on Eq.(1), 
flux density at frequency $\nu$ then declines 
as $S_{\nu} \propto t^{3 (1-p)/4}$, where $p$ is the index 
characterizing the power-law distribution of the shocked 
ISM electrons, $d n_{\rm e}' / d \gamma_{\rm e} \propto 
\gamma_{\rm e}^{-p}$. These expressions are valid only 
when $\gamma \gg 1$.

However, optical afterglows from GRB 970228 and GRB 970508 
were detected for as long as 190 and 260 d respectively, while 
in Eq.(1), even $t = 30$ d will lead to $\gamma \sim 1$. It is 
clear that the overall evolution of the postburst fireball can
not be regarded as a simple one-phase process, we should pay 
special attention to the transition from the ultra-relativistic 
phase to the non-relativistic phase.$^{10}$ This is unfortunately 
ignored in the literature. 

Second, the expansion of the fireball might be either adiabatic 
or highly radiative. Extensive attempts have been made to find 
a common model applicable for both cases.$^{11 - 13}$ As a result, 
a differential equation has been proposed by various authors,$^{12,13}$
\begin{equation}
\frac{d \gamma}{d m} = - \frac{\gamma^2 - 1} {M},
\end{equation}
where $m$ is the rest mass of the swept-up ISM, 
$M$ is the total mass in the co-moving frame, including 
internal energy $U$. Since thermal energy produced during the collisions 
is $ dE = c^2 (\gamma -1) dm $, usually we assume:$^{13}$ 
\begin{equation}
dM = { (1 - \epsilon) \over c^2} dE + dm 
    = [( 1 - \epsilon) \gamma + \epsilon ] dm, 
\end{equation}
where $\epsilon$ is defined as the fraction of the shock generated thermal 
energy (in the co-moving frame) that is radiated. 
It is putative that Eq.(2) is correct in both ultra-relativistic and 
non-relativistic phase, for both radiative and adiabatic fireballs. 

In the highly radiative case, 
$\epsilon = 1$, $dM = dm$, Eq.(2) reduces to, 
\begin{equation}
\frac{d\gamma}{dm} = - \frac{\gamma^2 - 1}{M_{\rm ej} + m}, 
\end{equation}
where $M_{\rm ej}$ is the mass ejected from the GRB central engine.  
Then an analytic solution is available,$^{11,13}$ which satisfies 
$\gamma \propto R^{-3}$ when $\gamma \gg 1$ and  
$v \propto R^{-3}$ when $\gamma \sim 1$, where $v$ is the bulk 
velocity of the material. These scaling laws indicate that Eq.(2) 
is really correct for highly radiative fireballs. 
In the adiabatic case, $\epsilon = 0$, $dM = \gamma dm$, Eq.(2) 
also has an analytic solution:$^{12}$ 
\begin{equation}
M = [ M_{\rm ej}^2 + 2 \gamma_0 M_{\rm ej} m + m^2]^{1/2},
\end{equation}
\begin{equation}
\gamma = \frac{m + \gamma_0 M_{\rm ej}}{M}, 
\end{equation}
where $\gamma_0$ is the initial value of $\gamma$. 
During the ultra-relativistic phase, 
Eqs.(5) and (6) do produce the familiar power-law 
$\gamma \propto R^{-3/2}$, which is often quoted for an adiabatic 
blastwave decelerating in a uniform medium. In the non-relativistic limit 
($\gamma \sim 1$, $m \gg \gamma_0 M_{\rm ej}$), Chiang and 
Dermer have derived $\gamma \approx 1 + \gamma_0 M_{\rm ej} / m$,$^{12}$  
so that they believe it also agrees with the Sedov solution, 
i.e., $v \propto R^{-3/2}$.$^{14}$
However we find that their approximation is not accurate,$^{15}$ because 
they have omitted some first-order infinitesimals of 
$\gamma_0 M_{\rm ej} / m$. The correct approximation could be obtained 
only by retaining all the first and second order infinitesimals, which in  
fact gives: $\gamma \approx 1 + (\gamma_0 M_{\rm ej}/m)^2 /2$, then we have   
$v \propto R^{-3}$.$^{15}$ This is not consistent with the Sedov solution! 
 
The problem is serious: (i) It means that the reliability
of Eq.(2) is questionable, although it does correctly  
reproduce the major features for radiative fireballs and even
for adiabatic fireballs in the ultra-relativistic limit. 
(ii) In the non-relativistic phase of the expansion, the fireball 
is more likely to be adiabatic rather than highly radiative. 
However, it is just in this condition that the conventional model fails. 
So any calculation made according to Eq.(2) will lead to serious 
deviations in the light curves during the non-relativistic phase. 

Third, for simplicity, it is usually assumed that $\epsilon$ is 
a constant during the expansion. But in realistic case this is not 
true. The fireball is expected to be highly radiative ($\epsilon = 1$) 
at first, due to significant synchrotron radiation. In only one 
or two days, it will evolve to an adiabatic one ($\epsilon = 0$) gradually. 
So $\epsilon$ should evolve with time.$^{16}$

Below, we will construct a new model that is no longer subject 
to the aforementioned problems.

In the fixed frame, since the total kinetic energy of the fireball is 
$E_{\rm K} = (\gamma - 1) (M_{\rm ej} + m) c^2 + (1 - \epsilon) 
\gamma U$,$^{17}$ and the radiated thermal energy 
is $\epsilon \gamma (\gamma - 1) dm \: c^2$,$^{11}$ we have: 
\begin{equation}
d[ (\gamma - 1)(M_{\rm ej} + m) c^2 + (1 - \epsilon) \gamma U] 
     = - \epsilon \gamma (\gamma - 1) c^2 dm. 
\end{equation}
For the item $U$, it is usually 
assumed: $d U =  c^2 (\gamma - 1) dm $.$^{17}$ 
Eq.(2) has been derived just in this way. 
However, the jump conditions\,$^{11}$ at the forward 
shock imply that $U = (\gamma - 1) m c^2 $, so we suggest that the correct 
expression for $dU$ should be: 
$dU = d[(\gamma - 1) m c^2] = (\gamma - 1) dm \: c^2 + m c^2 d \gamma$. 
Here we simply 
use $U = (\gamma - 1) m c^2$ and substitute it into Eq.(7), 
then it is easy to get:$^{15}$ 
\begin{equation}
\frac{d \gamma}{d m} = - \frac{\gamma^2 - 1}
       {M_{\rm ej} + \epsilon m + 2 ( 1 - \epsilon) \gamma m}. 
\end{equation}

In the highly radiative case ($\epsilon = 1$), Eq.(8) 
reduces to Eq.(4) exactly. While in the adiabatic case 
($\epsilon = 0$), Eq.(8) reduces to: 
\begin{equation}
\frac{d \gamma}{d m} = - \frac{\gamma^2 - 1}
       {M_{\rm ej} + 2 \gamma m}. 
\end{equation}
The analytic solution is: 
\begin{equation}
(\gamma - 1) M_{\rm ej} c^2 + (\gamma^2 - 1) m c^2 \equiv E_{\rm 0}.
\end{equation}
Then in the ultra-relativistic limit, we get the 
familiar relation of $\gamma \propto R^{-3/2}$, and in the 
non-relativistic limit, we get $v \propto R^{-3/2}$ as required 
by the Sedov solution.  From these analysises, we believe that 
Eq.(8) is really correct for both radiative and adiabatic 
fireballs, and in both ultra-relativistic and non-relativistic phase.

In realistic fireballs, $\epsilon$ is a variable dependent on the 
ratio of synchrotron-radiation-induced to expansion-induced loss 
rate of energy.$^{16}$ As usual, we assume that in the co-moving 
frame the magnetic field energy density is a 
fraction $\xi_{\rm B}^2$ of the energy density $e'$, 
$B'^2 /(8 \pi) = \xi_{\rm B}^2 e'$, and that the electron 
carries a fraction $\xi_{\rm e}$ of the energy, 
$\gamma_{\rm e,min} = \xi_{\rm e} (\gamma - 1) m_{\rm p} / m_{\rm e} 
+ 1 $, where $m_{\rm p}$ and $m_{\rm e}$ are proton and 
electron masses, respectively. The co-moving frame expansion time 
is $t'_{\rm ex} = R / (\gamma c)$, and the synchrotron cooling 
time is $t'_{\rm syn} = 6 \pi m_{\rm e} c / 
(\sigma_{\rm T} B'^2 \gamma_{\rm min,e})$, where $\sigma_{\rm T}$ 
is the Thompson cross section. Then we have:$^{16}$
\begin{equation}
\epsilon = \xi_{\rm e} \frac{t'^{-1}_{\rm syn}}
           {t'^{-1}_{\rm syn} + t'^{-1}_{\rm ex}}.
\end{equation}

We have evaluated Eqs.(8) and (11) numerically, 
bearing in mind that:$^{18}$  
\begin{equation}
dm = 4 \pi R^2 n m_{\rm p} dR,
\end{equation}
\begin{equation}
dR = v \gamma (\gamma + \sqrt{\gamma^2 -1}) dt. 
\end{equation}
We take $E_0 = 10^{52}$ erg, $n = 1$ cm$^{-3}$, 
$M_{\rm ej} = 2 \times 10^{-5}$ M$_{\odot}$. 
Figs.(1) and (2) illustrate 
the evolution of $\gamma$ and $R$, where full, dotted, and dashed 
lines correspond to constant $\epsilon$ values of 0, 0.5, 
and 1 respectively. Dash-dotted lines are plot by allowing 
$\epsilon$ to vary according to Eq.(11). 
It is clearly shown that our new  
model overcomes the shortcomings of Eq.(2).
For example, for highly radiative expansion, the dashed lines in these 
figures approximately satisfy $\gamma \propto t^{-3/7}$, 
$R \propto t^{1/7}$, $\gamma \propto R^{-3}$ 
when $\gamma \gg 1$, 
and $v \propto t^{-3/4}$, $R \propto t^{1/4}$, $v \propto R^{-3}$  
when $\gamma \sim 1$. While for adiabatic 
expansion, the full lines satisfy $\gamma \propto t^{-3/8}$, 
$R \propto t^{1/4}$, $\gamma \propto R^{-3/2}$ when $\gamma \gg 1$, 
and satisfy $v \propto t^{-3/5}$, $R \propto t^{2/5}$, 
$v \propto R^{-3/2}$ when $\gamma \sim 1$.

In order to compare with observations, we have also calculated the 
synchrotron radiation from the shocked ISM. Fig.(3) 
illustrates R band afterglows from the realistic fireball.  
We see that after entering the non-relativistic 
phase, the light curve becomes steeper only slightly, consistent 
with the prediction made by Wijers {\sl et al.}.$^7$ In contrast, 
Eq.(2) generally leads to a much sharper decline.$^{18}$ In our 
new model optical afterglows from GRB 970228 
are generally well fitted.

To conclude, 
current researches are mainly concentrated on the ultra-relativistic 
phase of the expansion of GRB remnants. The popular dynamic model 
is in fact incorrect for adiabatic fireballs during the 
non-relativistic phase. This is completely unnoticed in the 
literature. We have revised the model. Our new model has been 
shown to be correct in both ultra-relativistic and non-relativistic 
phase. The revision is of great importance, taking account of the 
following facts: (i) Optical afterglows lasting for more than 
100 -- 200 d have been observed  from some GRBs, the advent of 
the non-relativistic phase seems inevitable. (ii) Beaming effects 
also lead to a steepening in the optical light curve, 
non-relativistic effects should be considered carefully to tell 
whether GRB ejecta are beamed or not, which is crucial in 
understanding the GRB origin. (iii) HI supershells might be highly 
evolved GRB remnants,$^{19, 20}$ to address this question in 
detail, we should deal with non-relativistic blastwaves. Additionally 
we suggest that at very late stages, GRB remnants might become 
highly radiative again, just in the same way that 
supernova remnants do.$^{14}$ This might occur when the bulk 
velocity is just several tens kilometers per second.

\vspace{2.0cm}
\centerline{\large \bf Figure Caption}

\figcaption{Evolution of the bulk Lorentz factor $\gamma$. We take 
$E_0 = 10^{52}$ erg, $n=1$ cm$^{-3}$, $M_{\rm ej} = 2 \times 10^{-5}$ 
M$_{\odot}$. The full, dotted, and dashed lines correspond to 
$\epsilon = 0$ (adiabatic), 0.5 (partially radiative), and 1 (highly 
radiative) respectively. The dash-dotted line is plotted by 
allowing $\epsilon$ to evolve with time. 
\label{fig1}}

\vspace{0.3cm}
\figcaption{Evolution of the radius. Parameters and line
styles are the same as in Fig.1.
\label{fig2}}

\vspace{0.3cm}
\figcaption{R band afterglows from a realistic fireball. We take 
$p = 2.1$, $\xi_{\rm e} = 1.0$, $\xi_{\rm B}^2 = 0.01$. The distance 
$D$ is 18 Gpc. Other parameters are the same as in Fig.1. For the 
origin of the observational data points, please see Ref.[10].
\label{fig3}}


\begin{thebibliography}{}

\bibitem[]{} $^1$ G. J. Fishman and C. A. Meegan, Ann. Rev. Astron. 
  Astrophys. 33 (1995) 415.

\bibitem[]{} $^2$ T. P. Li and M. Wu, Chin. Phys. Lett. 14 (1997) 557.

\bibitem[]{} $^3$ E. Costa {\sl et al.}, Nature, 387 (1997) 783.

\bibitem[]{} $^4$ M. J. Rees and P. M\'{e}sz\'{a}ros, 
  Astrophys. J. 430 (1994) L93.  

\bibitem[]{} $^5$ R. Sari, R. Narayan and T. Piran, 
  Astrophys. J. 473 (1996) 204. 

\bibitem[]{} $^6$ E. Waxman, Astrophys. J. 485 (1997) L5.

\bibitem[]{} $^7$ R. A. M. J. Wijers, M. J. Rees and P. M\'{e}sz\'{a}ros,
  Mon. Not. R. Astron. Soc. 288 (1997) L51. 

\bibitem[]{} $^8$ R. Sari, Astrophys. J. 489 (1997) L37.

\bibitem[]{} $^9$ Y. F. Huang, Z. G. Dai and T. Lu, Chin. Phys. Lett. 
  15 (1998) 775.

\bibitem[]{} $^{10}$ Y. F. Huang, Z. G. Dai and T. Lu, 
  Astron. Astrophys. 336 (1998) L69.

\bibitem[]{} $^{11}$ R. D. Blandford and C. F. McKee,
  Phys. Fluids 19 (1976) 1130.

\bibitem[]{} $^{12}$ J. Chiang and C. D. Dermer,
  Astrophys. J. 512 (1999) 699.

\bibitem[]{} $^{13}$ T. Piran, Phys. Rep. 314 (1999) 575.

\bibitem[]{} $^{14}$ T. A. Lozinskaya, {\sl Supernovae and Stellar 
  Winds in the Interstellar Medium} (New York, AIP, 1992) Chap. 9.

\bibitem[]{} $^{15}$ Y. F. Huang, Z. G. Dai and T. Lu,
  Mon. Not. R. Astron. Soc. in press (1999).

\bibitem[]{} $^{16}$ Z. G. Dai, Y. F. Huang and T. Lu, 
  Astrophys. J. 520 (1999) 634.

\bibitem[]{} $^{17}$ A. Panaitescu, P. M\'{e}sz\'{a}ros and M. J. Rees,
  Astrophys. J. 503 (1998) 315.

\bibitem[]{} $^{18}$ Y. F. Huang, Z. G. Dai, D. M. Wei and T. Lu, 
  Mon. Not. R. Astron. Soc. 298 (1998) 459.  

\bibitem[]{} $^{19}$ A. Loeb and R. Perna, 
  Astrophys. J. 503 (1998) L35.

\bibitem[]{} $^{20}$ Y. N. Efremov, B. G. Elmegreen and 
  P. W. Hodge, Astrophys. J. 501 (1998) L35.

\end{thebibliography}
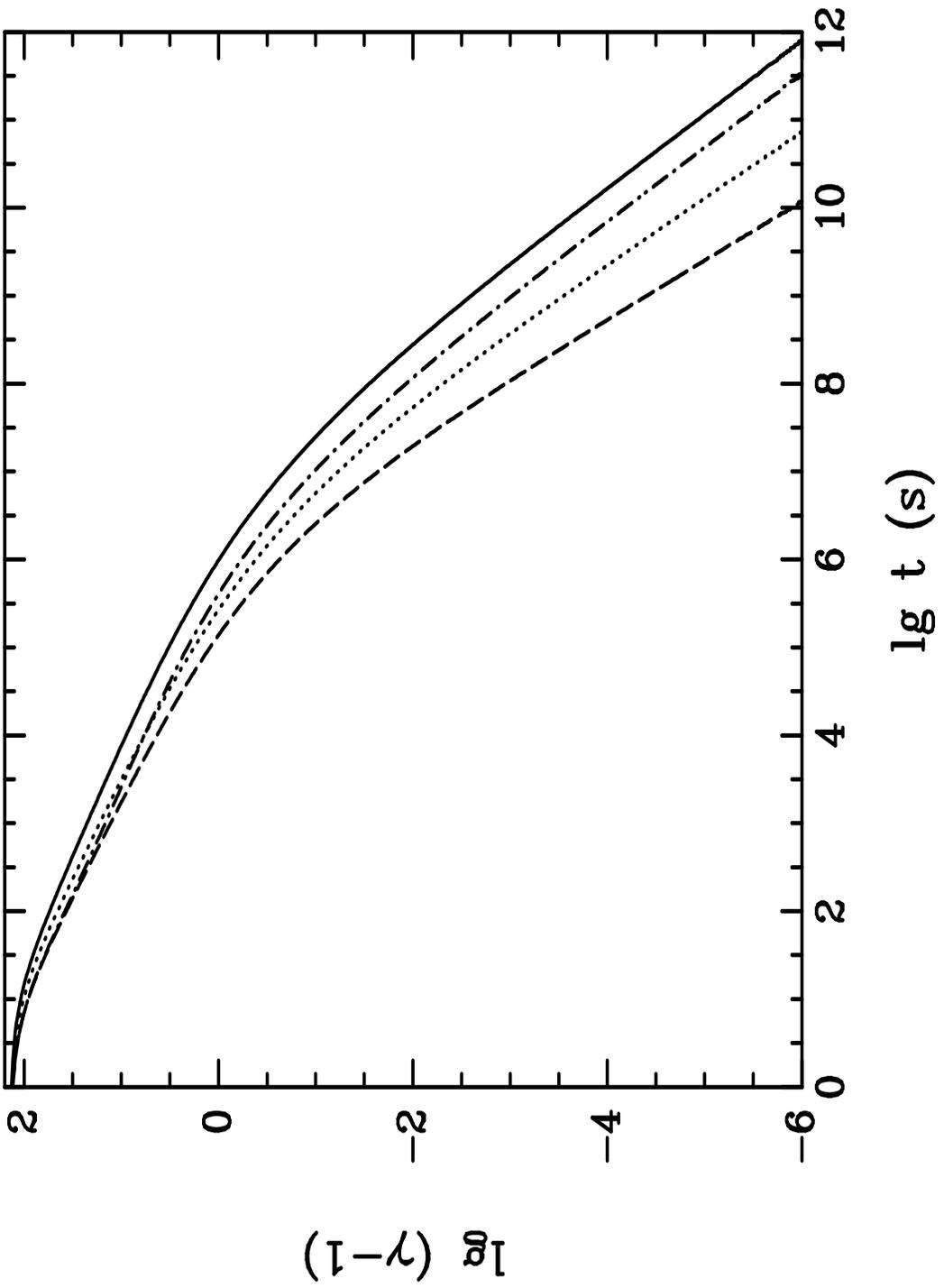
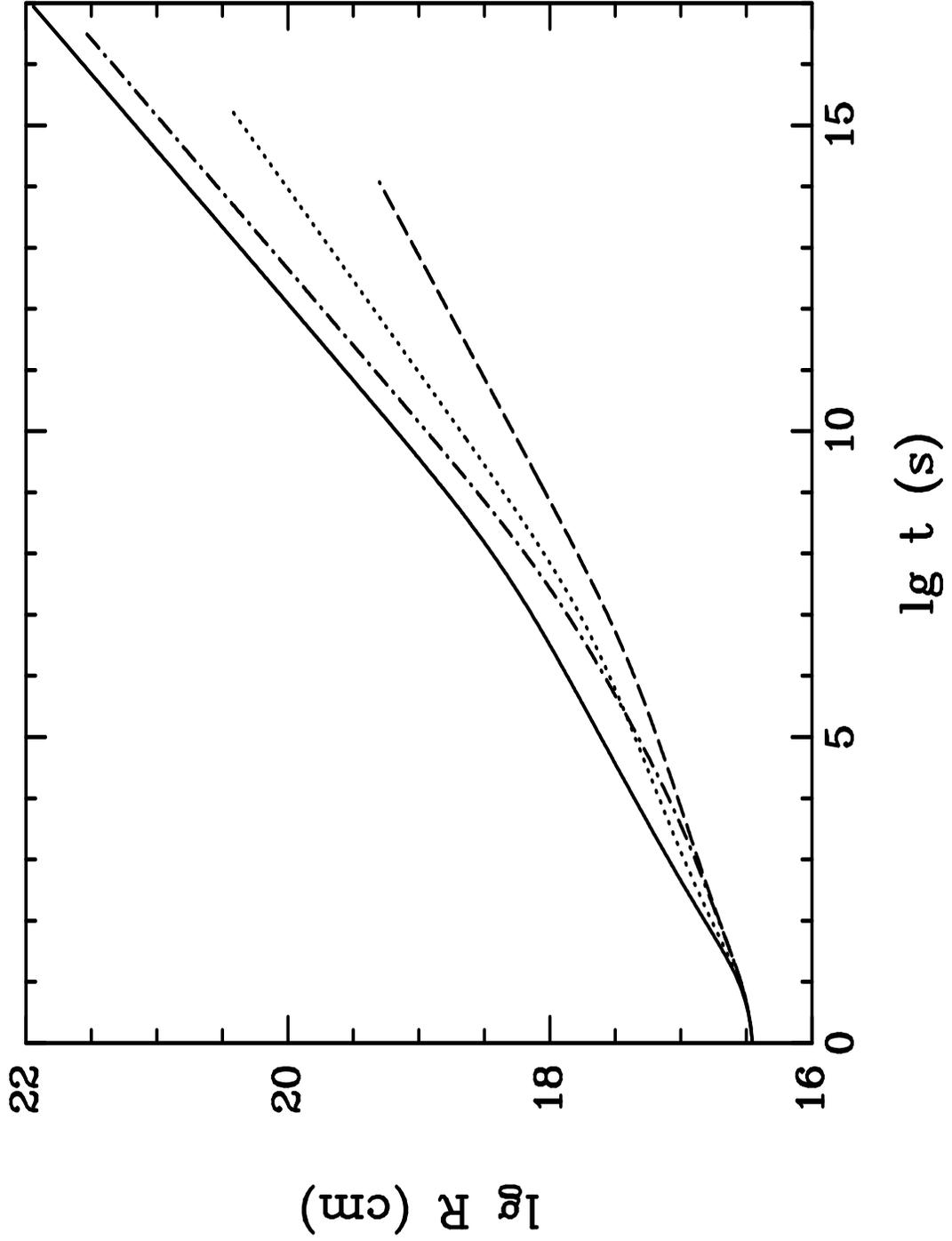
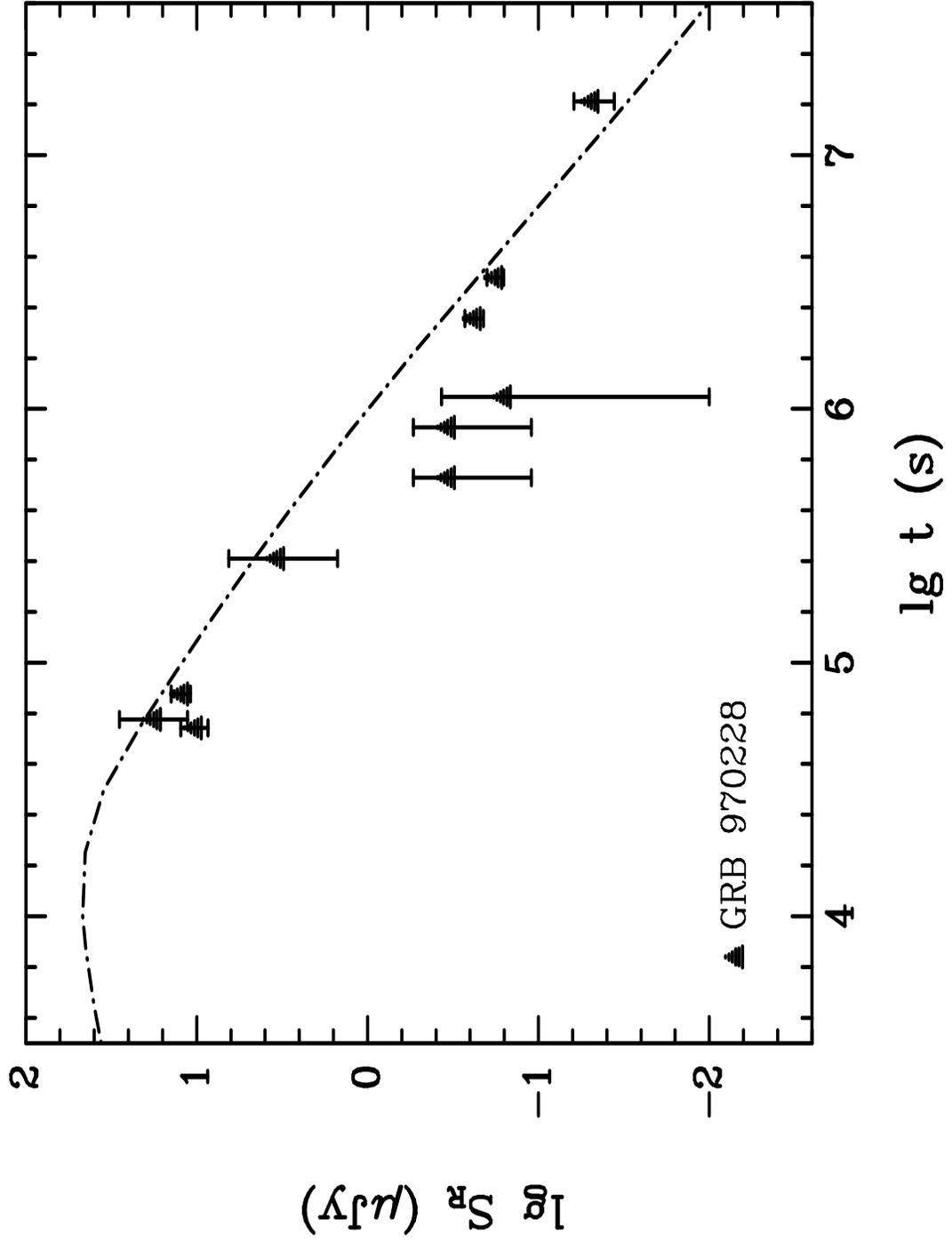
\end{document}